\documentclass[a4paper,twoside,10pt]{letter}
\usepackage{graphicx,saj,multicol,subeqnarray}

\def\papertype{{\doi DOI:10.2298/SAJ0673001P} \ \hfill\ Invited review}

\def\udc{524.7--82--48}
\begin{document}
\baselineskip=3.1truemm
\columnsep=.5truecm
\newenvironment{lefteqnarray}{\arraycolsep=0pt\begin{eqnarray}}
{\end{eqnarray}\protect\aftergroup\ignorespaces}
\newenvironment{lefteqnarray*}{\arraycolsep=0pt\begin{eqnarray*}}
{\end{eqnarray*}\protect\aftergroup\ignorespaces}
\newenvironment{leftsubeqnarray}{\arraycolsep=0pt\begin{subeqnarray}}
{\end{subeqnarray}\protect\aftergroup\ignorespaces}
%


\markboth{\eightrm THE BROAD LINE REGION OF AGN: KINEMATICS AND PHYSICS}
{\eightrm L.\v C. POPOVI\'C}

{\ }

\publ

\type

{\ }


\title{THE BROAD LINE REGION OF AGN: KINEMATICS AND PHYSICS}


\authors{L. \v C. Popovi\'c}

\vskip3mm


\address{Astronomical Observatory, Volgina 7, 11160 Belgrade 74, Serbia}


\dates{October 25, 2006}{October 25, 2006}


\summary{In this paper a discussion of kinematics and physics of the 
Broad Line Region (BLR) is given.  
The possible physical conditions in the BLR and problems in 
determination of 
the 
physical parameters (electron temperature and density) are considered. 
Moreover, one analyses the 
geometry of the BLR and the probability that (at 
least) a  fraction of the 
radiation in the Broad Emission Lines (BELs) originates from a relativistic 
accretion disk.}



\begin{multicols}{2}


\section{1. INTRODUCTION}

A group of galaxies emits intense permitted and forbidden lines from the
central part where a massive black hole is supposed to be located. This type
of galaxies,
so called Active Galactic Nuclei (AGN), are known as a most powerful sources of
radiation in the Universe. 

The investigation of nature of the emitting ionized gas in galactic nuclei is 
one of  important subjects in astrophysics today. Firstly, investigating  
the processes in the central parts of these objects we can learn about the 
innermost parts of other 'normal' galaxies. Secondly, AGN are the most powerful 
sources, located at different cosmological time-scales, and their 
investigation is cosmologically important. Finally, a part of emission from 
these objects  
(e.g. 
in the X-rays) has its origin very close to a massive black 
hole, and investigation of this emission can help us understand the 
physical processes in a strong gravitational field. 

The  narrow and broad emission lines  are present in spectra of AGN.
Their shapes and intensities give us opportunity to investigate the
physical and kinematic properties in the central part of AGN. Narrow
emission lines  originate from an extensive region (so called Narrow
Line Region - NLR) which can be resolved in the nearest AGN, while Broad
Emission Lines (BELs)  are formed in a very compact region (the so called Broad
Line
Region - BLR) in the central part of AGN (Osterbrock 1989, Krolik 1999, 
Peterson 2003). The BELs can be emitted by highly and lowly ionized emitters (so 
called High Ionization Lines - HILs and Low Ionization Lines - LILs).  
The investigation of BEL
shapes provide information about  conditions of the emitting gas
surrounding a massive black hole, assumed to be in the center of  these objects.

Indirect techniques such as reverberation mapping
(e.g. Peterson 1993) can, in principle, provide
information concerning the spatial distribution of the
emitting gas, but in practice require long monitoring
campaigns and are often inferred by  technical and
interpretation difficulties (e.g. Peterson et al. 1999). Whilst these studies 
appear to
favor Keplerian dynamics, it is unclear whether or
not the gas distribution has axial or spherical
symmetry (e.g.  Kaspi et al. 2000).
Spectropolarimetry provides an alternative
approach to investigate the nature of the BLR (see e.g. 
Smith et al. 2005). The
unique diagnostic strength of this technique is that
the polarization state of scattered light carries the
imprint of the scattering geometry, allowing the
structure and kinematics of both the scattering
medium and the emission source to be investigated
in unresolved sources. Moreover,  spectropolarimetry 
provides us an evidence that in some Sy 2 galaxies  the BLR is present (e.g. in the 
case of Mrk 533, see Miller and Gooldrich
1990).

Besides the strong emission lines, in spectra  of the 
AGN with BELs, the Broad Absorption Lines (BAL) in the UV part of spectra are 
present. Approximately 10\% of all quasars are with broad, blue-shifted
absorption lines. The outflow velocity can reach 0.1-0.2 c. Usually, in
their spectra the high ionization species as C IV $\lambda$1549, Si IV
$\lambda$1397, N V
$\lambda$1240 and Ly$\alpha$ lines have been observed. Rarely some of them
also exhibit broad absorption lines of Mg II $\lambda$2798 and Al III
$\lambda$1857, low ionization lines. Broad
absorption lines may have different shapes, and also differences in
continua are present in spectra of various types of these objects
(see e.g. Reichard et  al. 2003).
 The spectrum of a Broad Absorption Line Quasar (BALQSO) is usually
interpreted as a superposition of the continuum emission from the central engine with 
broad emission lines from the BLR  created
near the center of a QSO and the broad absorption lines
emitted from a separate outlying region - Broad Absorption Line
Region (BALR).   

In at least four last decades, a large  number of papers  about the BLR 
structure has been published (see e.g. review of Sulentic et al. 2000, Dultzin-Hacyan 
et al. 2000); also, a unified 
model of all AGN was proposed and discussed (e.g. Antonucci 1993, Elvis 
2000, L\'ipari and Terlevich 2006). But some 
unanswered questions concerning the BLR are still present:

(i) {\bf Physics of the BLR}. Since the BLR is located on a relatively small 
distances from the extremely powerful 
energy source of AGN, the matter in the BLR
 is likely to be in a physical environment which can hardly be compared to
that in other well studied
 astrophysical objects. It seems that plasma in the BLR is in a condition that is 
closer to stellar atmospheres than to photoionized nebulae (Osterbrock 1989). As a 
direct consequence, many of the custom any techniques, that have been
 derived to identify the physical properties of photoionized nebulae, 
are often
 unable to provide reliable answers or even to be applied in the case of the 
BLR.
Some approximation methods can be applied to probe the physics of the BLR, but 
they are still far
from taking us to a
 detailed solution. Also, the connection between BALR and BLR is not clear. Very 
broad 
absorption lines in the UV spectra of
AGN indicate that they should originate close to the AGN engine, but it is not
clear yet where this region is placed, closer to or further away from the BLR, or even a 
part of the BLR is in such a condition that it is able to absorb radiation in the UV. 

(ii) {\bf  Kinematics of the BLR}. The strong gravitation field is often 
taken into account
in order to constrain the geometry of the BLR, but the true BLR geometry is not yet 
clearly known. There is a small fraction of AGN with double peaked 
broad lines which indicate presence of an accretion disk in the BLR, but the number
of such objects is statistically insignificant (around 5\%) for conclusion 
about the disk presence. On the other hand, various geometries can be considered 
(spherically distributed 
clouds, jets, etc.). Also, it is not  clear if the BLR is composed of more than 
one geometrically
consistent region, or it is a combination of two or even more geometrically 
different regions (e.g. disk+jets, or spherical region+disk, etc.).

In this review,  we will give some ideas about physics and kinematics of the 
BLR. 
The aim of the paper
is to give 
an overview of the 
investigation of the BLR physics and kinematics, especially the investigation
 that were carried out by our 
group.

\vskip-2mm

\section{2. BROAD LINE REGION: KINEMATICS}

\vskip-1mm

Various geometries can be assumed in describing the BLR (see Sulentic 
et al. 2000 in more details). The BLR geometry (kinematics) affects the broad 
line shapes. We should note here that due to  high random velocities (about 
several thousand km/s) we can expect that Doppler effect is dominant among other 
broadening mechanisms (broadening due to collisions, natural broadening, 
etc.). Therefore, in order to grasp the BLR kinematics, in first approximation 
one  can fit the broad lines with a number of Gaussin functions (see Figs. 1 
and 2). First step in the investigation of the BEL shapes is to clean a BEL 
from the narrow and satellite lines. 

The rotating  accretion disk model (see e.g. 
Perez et 
al. 1988, Chen et al. 1989, Chen and Halpern 1989, Dumont and 
Collin-Suffrin 1990ab, Dumont et al. 1991, Dumont and Joly 1992,
Eracleous 
and Halpern 1994, 2003, Sulentic et al. 1998,  Pariev and Bromley 1998, 
Rokaki and 
Boisson 1999, Shapovalova 
et 
al. 2004, 
Popovi\'c  et al. 2001a, 2002, 2003a, 2004, Kollatschny and Bischoff 2002, 
Bon et al. 
2006) has been very 
often 
discussed in 
order to explain the observed broad optical emission-line profiles in AGN. This model 
fits well the widely accepted AGN paradigm  that the "central engine" consists of a 
massive black hole fueled by an accretion disk. 
 However, the
fraction of AGN clearly showing double-peaked profiles is small
 and   statistically insignificant (e.g. Strateva et al. 2003).
On one hand, the presence of double-peaked lines is not
required as a necessary  condition for the existence of a disk
geometry in BLRs. Even if the emission in a spectral line comes
from a disk, the parameters of the disk (e.g. the inclination) can be
such that one observes single-peaked lines
(e.g. Popovi\'c et al. 2004, Ili\'c et al. 2006). Also, a Keplerian disk with
disk wind can produce single-peaked broad emission lines as
normally seen in most of the AGN (Murray and Chang 1997). On the other
hand, taking into account the complexity of emission line regions
of AGN (see e.g. Sulentic et al. 2000), one might expect that the broad
emission lines are composed of radiation from two or more
kinematically and physically different emission regions, i.e. that
multiple BLR emission components with fundamentally different
velocity distributions are present (see e.g. Romano et al. 1996).
Consequently, one possibility could be that the emission of the
disk is masked by the emission of another emission line region.
Recently, in several papers (Popovi\'c et al. 2001a, 2002, 2003a, 2004, Bon et al. 
2006, Ili\'c et al. 2006) the 
possibility that the disk emission is present in the AGN having single peaked 
lines is investigated
and one can conclude that it is likely that the disk emission mainly contributes 
to the wings 
of BELs.
 This supports the idea that the broad
optical lines  originate in more than one emission region, i.e.
that the Broad Line Emission Region is complex and composed of at
least two regions (e.g. Ili\'c et al. 2006). Note here that Corbin and Boroson 
(1996) found that 'the
difference between the Ly$\alpha$ and H$\beta$ full width at zero
intensity (FWZI) values provides additional evidence of an
optically thin very broad line region (VBLR) lying inside an
intermediate line region (ILR) producing the profile cores'.
Consequently, one may expect that the VBLR can be formed in a disk
or disk-like emission region. 

\subsection{2.1 Two-component model of the BLR}

As we mentioned,
Corbin and Borson (1996) investigated the combined ultraviolet and optical
spectra of 48 QSOs and Seyfert 1 galaxies in the redshift range
0.034-0.774. They found a statistically significant difference
between the FWZI distributions of the Ly$\alpha$ and H$\beta$
lines. The difference between the Ly$\alpha$ and H$\beta$ FWZI
values provides additional evidence for an optically thin VBLR
(which might be a disk or disk-like region) which contributes to
the line wings. It is located inside an ILR which produces the
profile cores. Also, they found a { relatively weak 
correlation between the UV profile asymmetries and widths and
those of the H$\beta$ line}. This suggests a stratified  structure of
the BLR, consistent with the variability studies of Seyfert 1
galaxies (see e.g. Kollatschny 2003). The smaller average FWHM
values
of the UV lines compared to the H$\beta$ indicate that the ILR emission makes a
higher contribution to the UV lines,
whereas in the Balmer lines the VBLR component is more dominant. This is
also the case
in well known AGN with double-peaked Balmer lines, which usually
show a single-peaked  Ly$\alpha$ line (see e.g. the case of Arp
102B, Halpern et al. 1996).

The wings of the broad H$\alpha$ emission line in the spectra of a
large sample of AGN (around 100 spectra) were also investigated by
Romano et al. (1996). They found an indication of multiple BLR emission.
Although a two-component model can probably be represented by
other geometries, we choose the one with a disk giving the wings
of the lines, and a spherical component giving the core of the
line.
We assume that the kinematics of the additional
emission
region can be described as the emission of a spherical  region
with
{ an isotropic velocity distribution}, i.e. with
a local broadening $w_G$ and shift $z_G$. Consequently, the emission
line
profile
can be described by a Gaussian function.
The whole line profile can be described by the relation (Popovi\'c et al. 2004):

$$I(\lambda)=I_{AD}(\lambda)+I_G(\lambda)$$
where $I_{AD}(\lambda)$,  $I_G(\lambda)$ are the  emissions of the
relativistic accretion disk and of  an additional  region,
respectively. The model capable reproducing broad line profiles (see Fig. 3), 
but the problem is the number of needed parametes which cannot be constrained properly 
(see discussion in Popovi\'c et al. 2004 and Bon et al. 2006).

We should mention that besides a disk (or a disk-like region) or spiral shock 
waves within a disk (Chen et al. 1989, Chen and Halpern 1989), other 
geometries may cause the same kinds of substructure 

\centerline{\includegraphics[height=8cm]{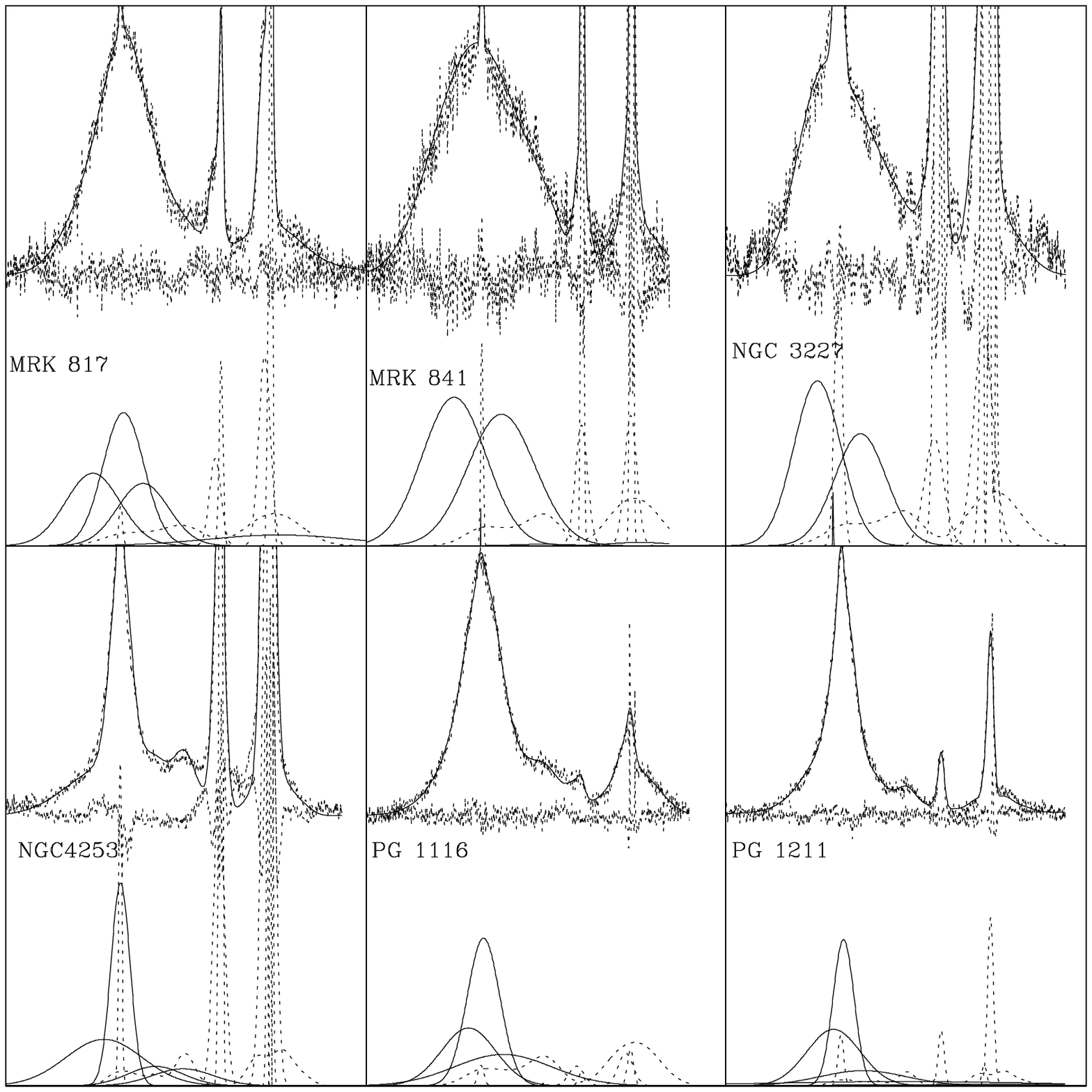}}

\vspace{-.5cm}

\figurecaption{1.}{Gaussian multicomponent analysis of the  emission lines
indicates
very complex kinematics of the BLR and NLR (Popovi\'c et al. 2004).}

\centerline{\includegraphics[width=8cm]{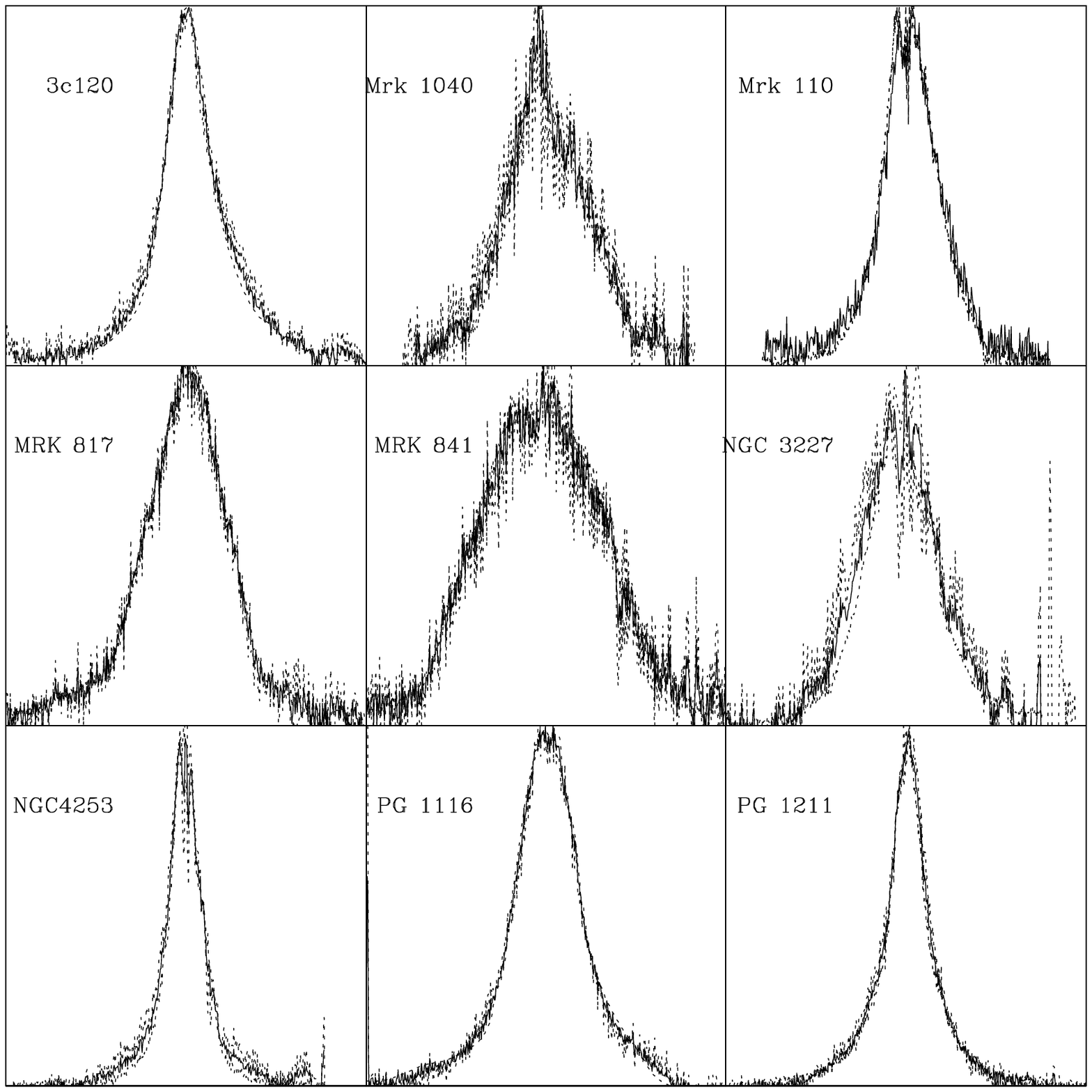}}

\vspace{-.5cm}

\figurecaption{2.} {The broad line profiles after subtraction of the narrow and 
satellite lines
(Popovi\'c et al. 2004).}

\centerline{\includegraphics[height=8cm]{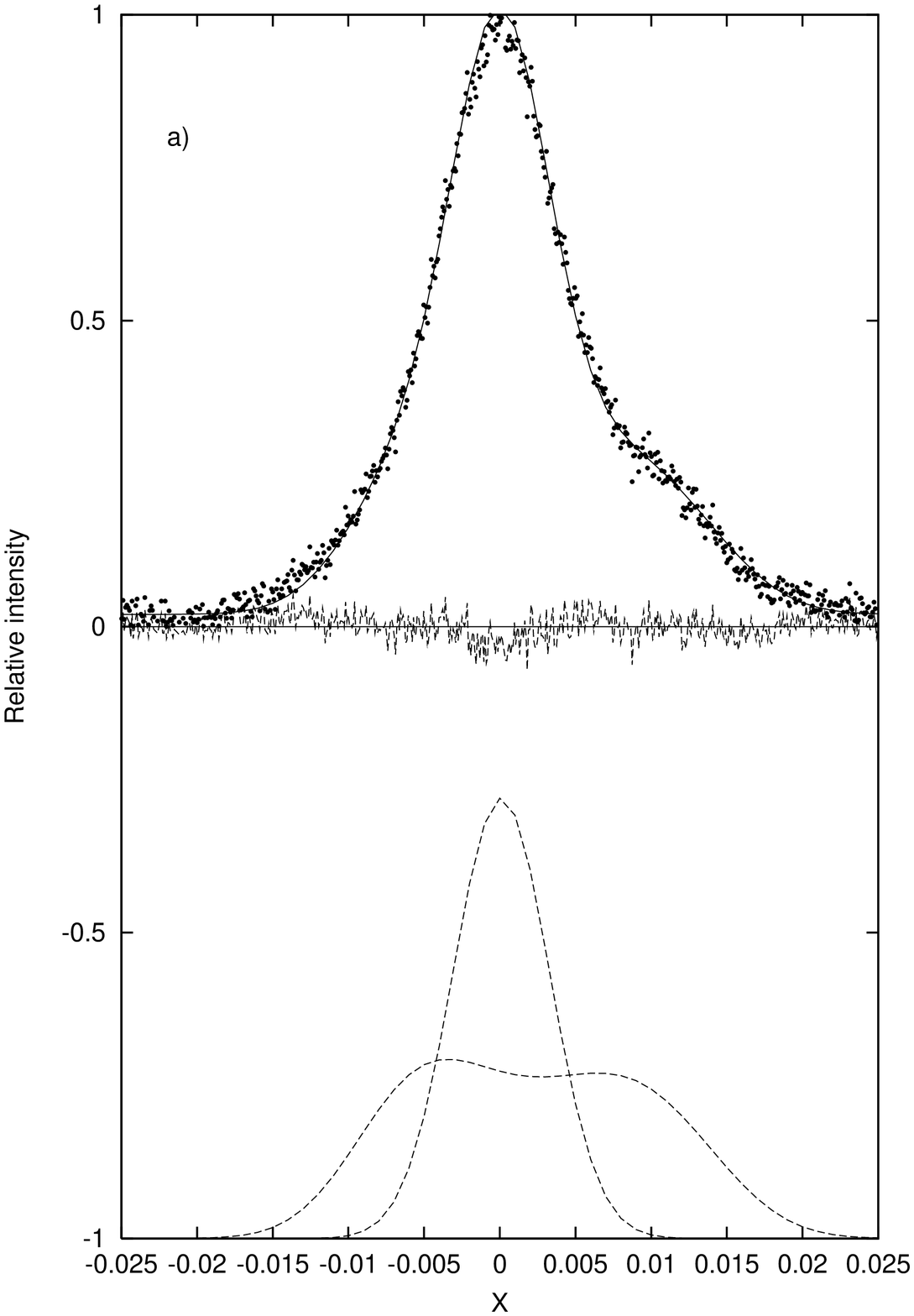}}
\figurecaption{3.}{Two-component model fit of the broad emission line: below 
are the components corresponding to the line core and wings. The double 
peaked component that predominantly contributes to the line wings is assumed 
to be emitted from an accretion disk (Popovi\'c et al. 2004).} 

\noindent in the line
profiles: i) emission from the oppositely-directed sides of a
bipolar outflow (Zheng et al. 1990, 1991); ii) emission from a
spherical system of clouds in randomly inclined Keplerian orbits
illuminated anisotropically from the center (Goad and Wanders 1996); or iii)
emission from a binary black hole system (Gaskell 1982, Gaskell 1996, Popovi\'c et al. 2000). In
any case, one should consider a two-component model with an ILR
contributing to the broad line cores  and one additional emitting
region contributing to the broad line wings. Recent investigations
(see e.g. Wang et al. 2003, Eracleous and Halpern 2003)
 have shown that the disk geometry for VBLR may be accepted as a
reality. Moreover, Eracleous and Halpern (2003) found  that  the disk emission is
more successful  not only in explaining double-peaked line
profiles but also in interpreting   another spectroscopic
properties of AGN presenting these  double-peaked Balmer lines.

\subsection{2.2 Investigation the BLR geometry by gravitational microlensing}

Gravitational lensing is in general achromatic: the deflection angle of a
light ray does not depend on its wavelength. However, the
wavelength-dependent geometry of the various emission regions may
result in chromatic effects (see
Popovi\'c and Chartas 2005, and references therein). Studies aimed at
determining the
influence
of microlensing on  spectra of lensed quasars (hereafter QSOs)
ought to account for the
complex structure of the QSO central emitting region. Since the sizes of
the emitting regions are wavelength-dependent, microlensing by stars in
a lens galaxy will lead to a wavelength-dependent magnification. The
geometries of the line and the continuum emission regions are, in general,
different and there may exist a variety of geometries depending on the type
of AGN (i.e. spherical, disc-like, cylindrical, etc.). Observations and
modeling of microlensing of the BLR of lensed QSOs
are promising, because the study of
the variations of the BEL shapes in a microlensed QSO
image could constrain the size of the BLR and the continuum region.

Continuum-line reverberation experiments with low-redshift QSOs
tell us that the broad-emission line region (BLR) is significantly
smaller than earlier assumed, and it is typically several light
days up to a light year across (e.g., Kaspi et al.\ 2000). It
means that the BLR radiation could be  significantly amplified due to
microlensing  by  (star-size) objects in an intervening galaxies
(Abajas et al. 2002, 2005).
Hence, gravitational lensing can provide an additional method for
studying the inner structure of high-redshift quasars for several
reasons:

(i) the extra flux magnification, from a few to 100 times,
due to the lensing effect enables us to obtain high
signal-to-noise ratio (S/N) spectra of distant (high-redshifted)
 quasars with less
observing time;

(ii) the magnification of the spectra of the different images may
be chromatic (e.g.
Popovi\'c and
Chartas 2005, Popovi\'c et al. 2006ab, 
Jovanovi\'c 2006) since the spectral line and the continuum
emitting regions are different in sizes and geometrically complex
and/or complex gravitational potential of lensing galaxy; (iia)
consequently, microlensing events lead to wavelength-dependent
magnifications of the continuum that can be used as indicators of
their presence (Popovi\'c and Chartas 2005);

(iii) gravitational microlensing can also change the shape of the
broad lines (see Popovi\'c et al. 2001b,c, Abajas et al. 2002, 2005,
Popovi\'c and Chartas 2005),  the deviation of the line profile
depends on the geometry of the BLR.

Finally, the monitoring  of lensed QSOs in order  to investigate
the effect of lensing on the spectra  can be useful
not only for constraining the unresolved structure of the central
regions of QSOs, but also for providing insight into the complex
structure of the lens galaxy.

Taking into account the redshift of lensed QSOs, one ought to
obtain the spectra from 3500 to 9000 \ \AA\  (covering also, the
broad C IV, CIII and Mg II lines which are emitted from the BLR
region) of a sample multi-imaged QSOs. Concerning the estimation of the
BLR dimensions (see Kaspi et al. 2000), one should select a sample of
lensed QSOs  where  the BLR  microlensing might be expected (Abajas et al. 2002).
  To find the
possible microlensing one can apply the method given by
Popovi\'c and Chartas (2005) comparing the spectra (in the
continuum and in the broad lines) of different components  in order
to detect the difference caused by microlensing or/and
millilensing. Using previous theoretical estimates of

\end{multicols}

\vfill\eject

\begin{multicols}{2}

\noindent line shape
variations due to microlensing (Popovi\'c et al. 2001bc,
Abajas et. al. 2002, 2005, Popovi\'c and Chartas 
2005, Popovi\'c et al. 2003a,b, 2006ab, see 
also Fig. 4),  the observed
spectra can
be fitted to the theoretical line profile, assuming different geometries.
Thence, one will be able to  estimate the geometry and dimension
of the BLR. Also, comparing difference in amplifications, of the
continuum, C IV and Mg II lines one will be able to conclude about
differences between high and low ionized line emitting regions and
compare them with the size and geometry of the continuum emission
region.

{\ }

\centerline{\includegraphics[width=8cm]{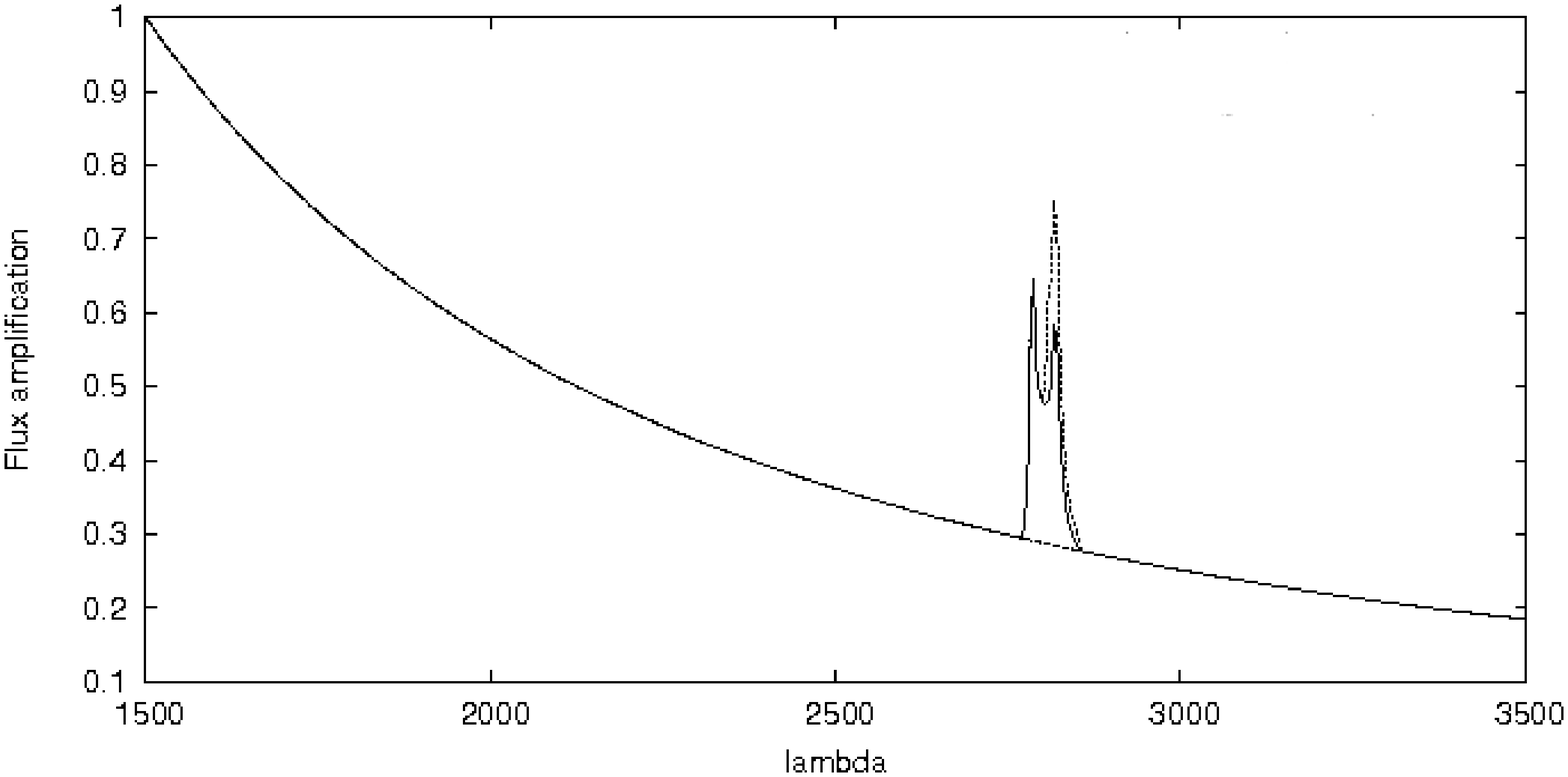}}
\centerline{\includegraphics[width=8cm]{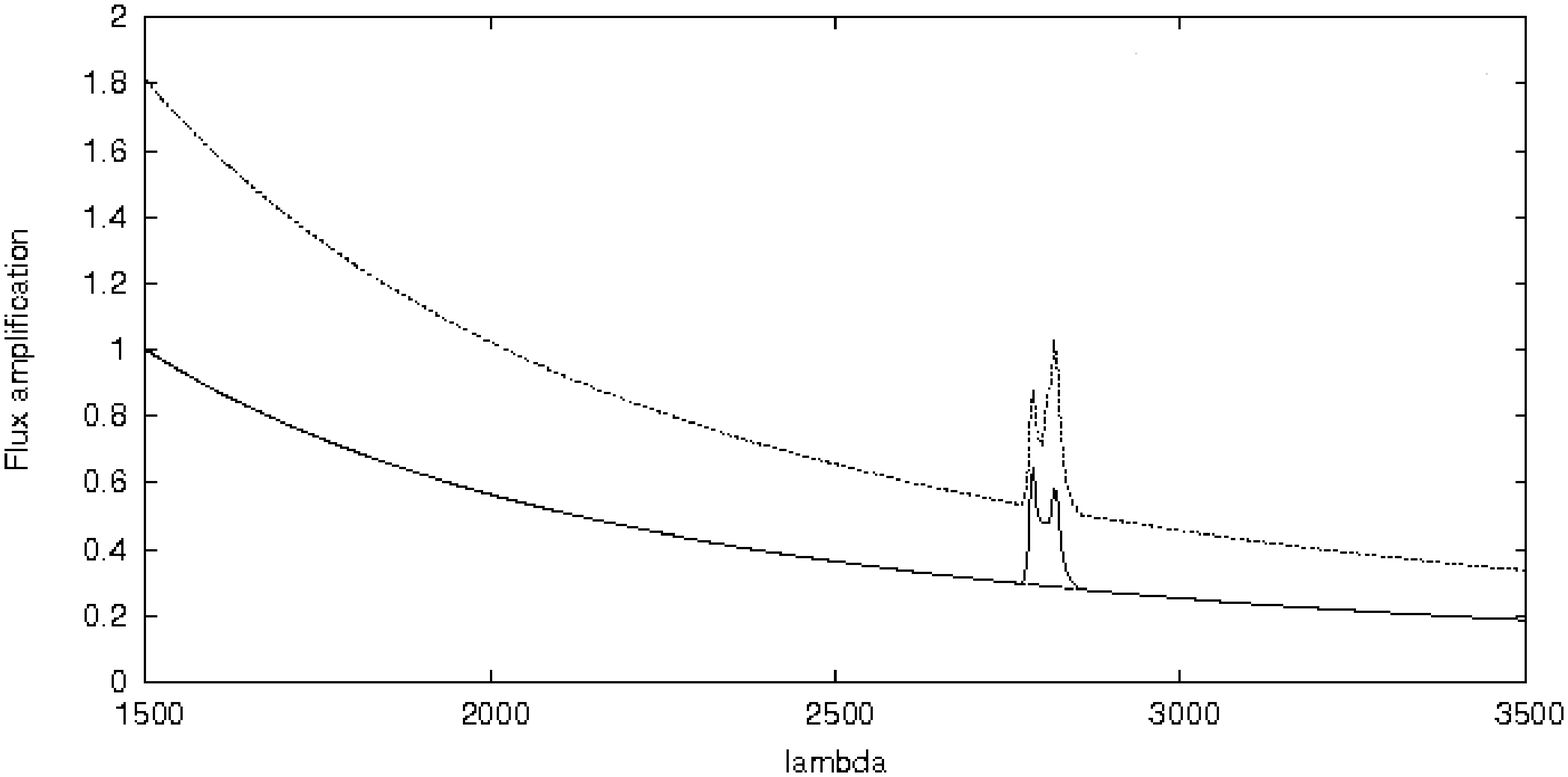}}
\figurecaption{4.}{The simulation of the variation in Mg II, $\lambda=2798$\AA\
and the continuum between 1500 \AA\ and 3500 \AA\ due to microlensing by
straight fold caustic. The microlensing may affect only line profile (top)
as well as the line profile and the continuum (down). Amplified spectra is presented by dashed line
(Popovi\'c 
2005).}

As an example, let us discuss the variability  in the shape of Mg II line of
images A and B of J0924+021 observed in two epochs (on 14/01/2005 and
01/02/2005) 
by Eigenbrod et al. (2006). 

Comparing
the Mg II line shapes of components A and B, we 
concluded that the shape of
Mg II was changed in the A image observed on 01/02/2005. There is no
significant difference between line profiles of  Mg II lines observed on
14/01/2005 between  A and B components (only they are amplified by
different factor due to lensing effect). Also, there is no significant
 difference
between line profiles of the
component B observed on  01/02/2005 and 14/01/2005.
 We 
used test described in Popovi\'c and Chartas (2005), and found that it was 
probably 
microlensing that caused this variation, 
as it was noted also in Eigenbroad et al. (2006).
To explain this, we apply the two component model, assuming that only the 
disk is 
microlensed (due to microlening with Einstein Radius Ring at about  
several thousand 
gravitational radii). 
As one can see 
in Fig. 5, the model can explain registred changes in the line profile from the two 
epochs. Note here that the similar variation in the line profiles of lensed QSO SDSS J1004+4112 was reported by
 Richards et al.  (2004), Lamer et al. (2006) and G\'omez-\'Alvarez et al. (2006). 

{\ }

\centerline{\includegraphics[width=8cm]{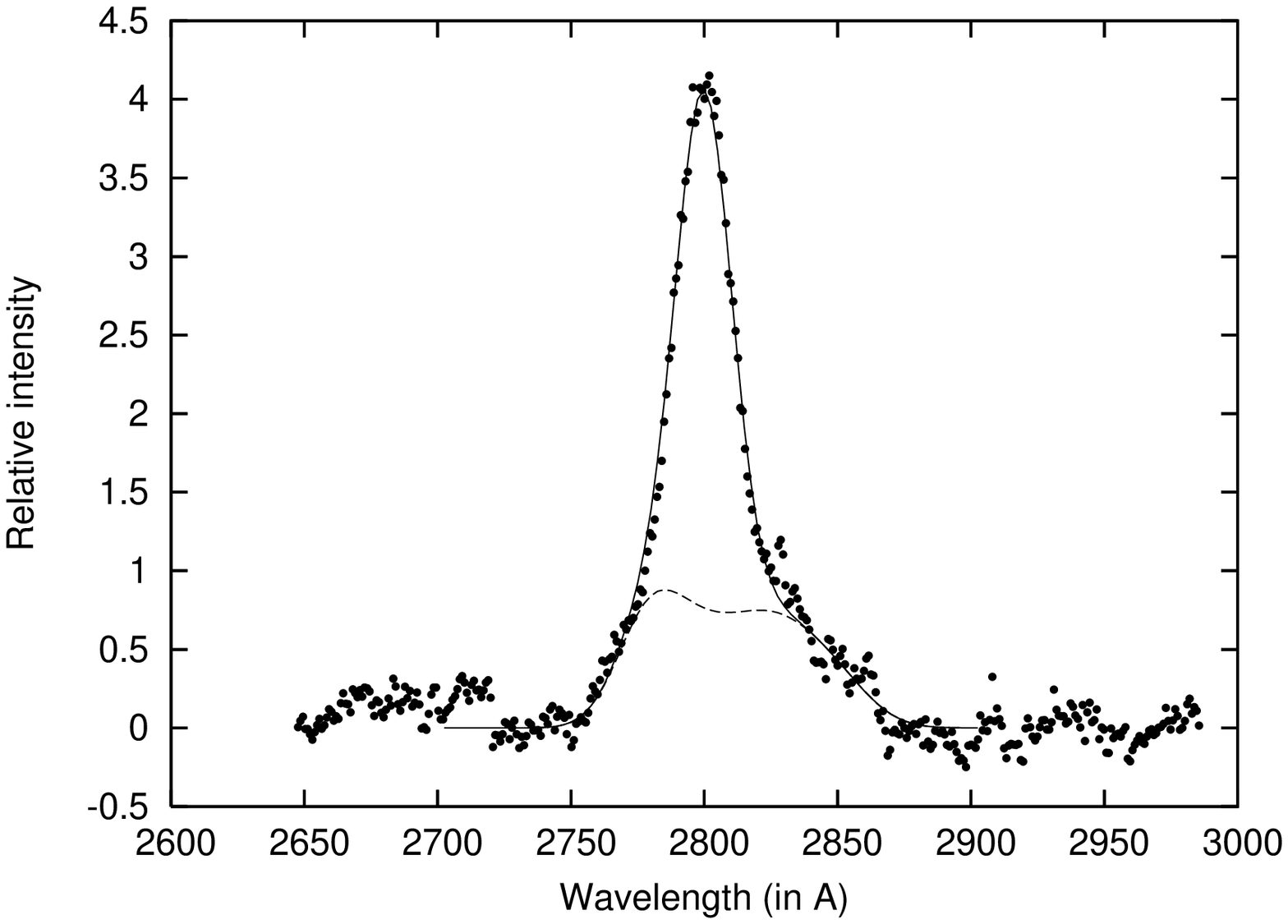}}
\centerline{\includegraphics[width=8cm]{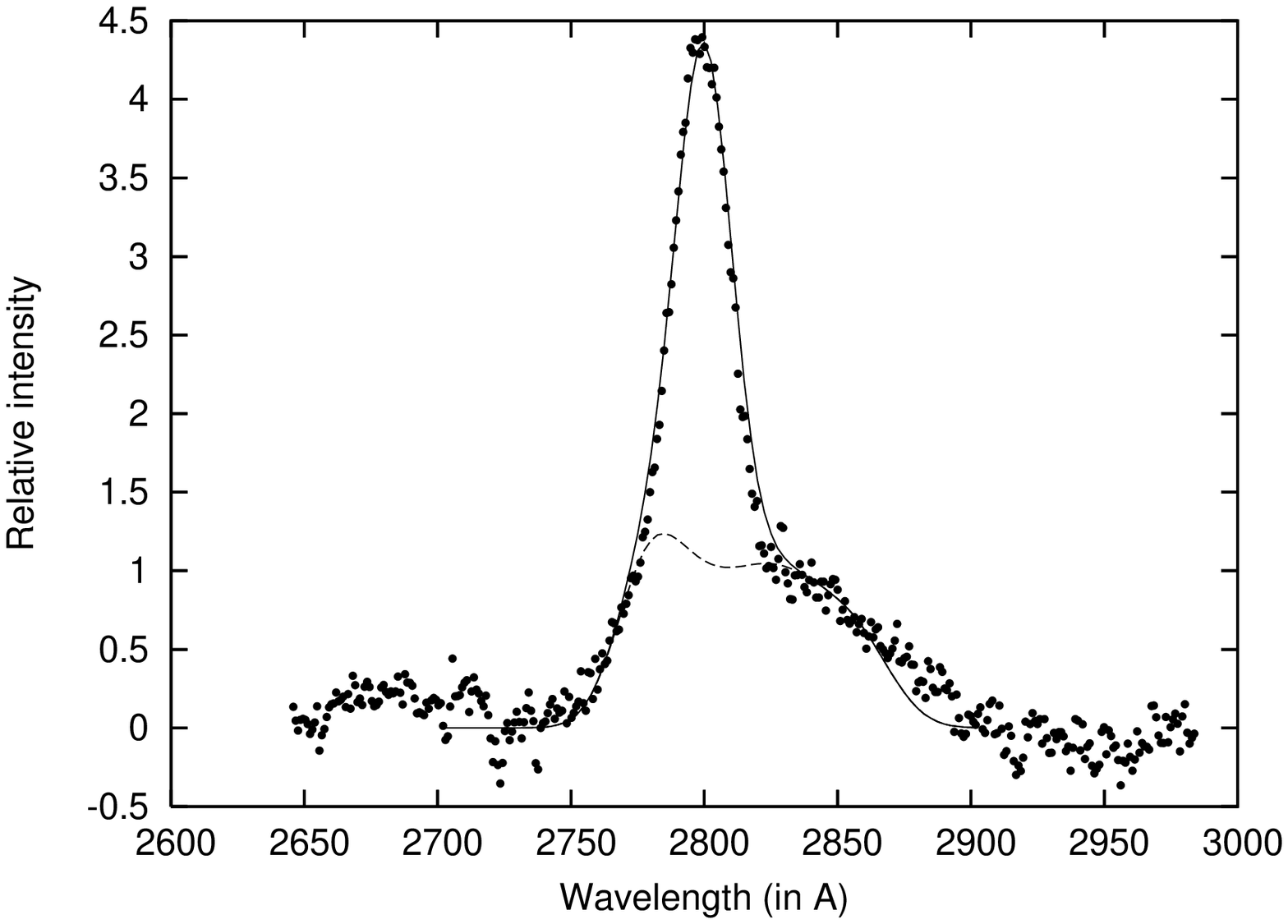}}
\centerline{\includegraphics[width=8cm]{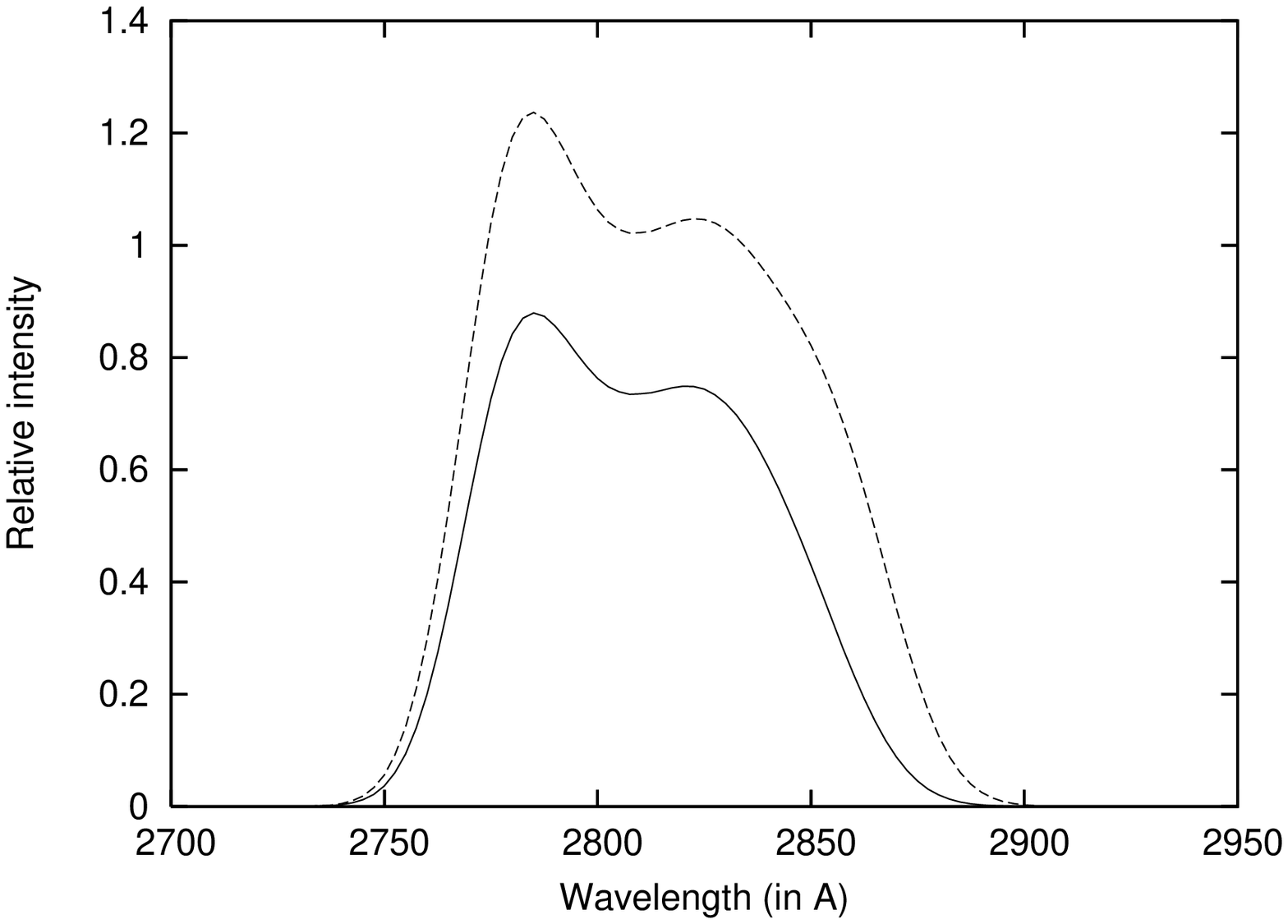}}
\figurecaption{5.}{The best fit with the two-component model non-amplified line
  (up) and after including microlensing of the disk, amplified line (middle).
Down is shown the comparison the non-amplified disk emission line  (solid line)
  with amplified (dashed line).}



\section{3. PHYSICS OF THE BLR: DIAGNOSTIC BY USING EMISSION LINES}

Physics and kinematics in the Broad Line Region are more
complicated than in the Narrow Line Region (NLR) or in gaseous nebulae
(Osterbrock 1989, Krolik 1999, Sulentic et al. 2000 and
references  therein). In { contrast to
 the NLR} where forbidden lines (e.g. [OIII] and [NII] lines\footnote{These lines also can be used to check 
sophisticated calculation of atomic parameters, see e.g. Dimitrijevi\'c et al. (2006)}) can be
used { as} emitting plasma diagnostics,  the physical conditions in
the  BLR cannot be understood using simple relations between the line
ratios.   The pure recombination conditions cannot be applied in the BLRs,
e.g. the flux line ratios are different from those expected in the case of
recombination (e.g. in some AGN $Ly\alpha$/H$\beta \approx$ 10,
Osterbrock 1989).
Moreover, the optical spectrum of many AGN is dominated by two broad permitted Fe 
II emission 
line
blends, one centred at about 4570 \AA\  and the other centred at 5350 \AA . 
Optical
Fe II has now been measured in the spectrum of several hundred broad-line AGN, 
showing a
large range of intensity relative to Balmer recombination lines.
From the theoretical point of view, considerable efforts have been devoted to
understanding the origin of Fe II optical emission in broad-line AGN during the 
last few
decades. However, extreme Fe II emission is not well explained by standard
photoionization models. There are several ideas about the
 source of heating for the Fe II emitting gas (as e.g. zones of the BLR of very 
high  opacity to ionizing radiation, see Kwan and Krolik 1981; jets responsible for 
the  compact radio sources, see  Norman and Miley 1984), but one can also indicate 
collisional 
ionization process (Joly 1988, 1989, 1991, V\'eron-Cetty et al. 2006), which 
could be
related mainly to shocks.

{Several}
effects can result in
such a ratio of the line flux of hydrogen (and another AGN) lines\footnote{In 
the last over
30 years, there appeared numerous papers devoted this problem, see e.g.
  Netzer (1975, 1976), Ferland and Netzer (1979),  Ferland et al. (1979), Kwan
(1984), Kallman and Krolik (1986), Collin-Souffrin (1986), Collin-Souffrin and Dumont 
(1986),
Rees et al. (1989), Ferland et al. (1992), Shields and Ferland (1993),
Dumont et
al. (1998), etc.} {between them, also the} collisional excitation
and
extinction effects.
Dust is present in the { host galaxy of an AGN
(see e.g. Crenshaw et al. 2002, Crenshaw et al. 2004, Gabel et al. 2005,
etc.)}, but it seems that
in some cases it cannot explain the line flux ratios.
{ Also, in the BLR different mechanisms capable of affecting the
spectral lines can be present. The classical  studies point toward
photoionization
as the main heating source for  the BLR emitting gas (see e.g. Kwan
and Krolik 1981,  Osterbrock 1989, Baldwin et al. 1995, Marziani et al.
1996, Baldwin et al. 1996, Ferland et al. 1998, Krolik 1999, Korista and
Goad 2004) that may explain observed spectra of AGN. But, in some
cases as e.g.  in Dumont et al. (1998) and V\'eron-Cetty et al. 2006, 
a favor is given to  
a non-radiatively heated region that contributes to the BLR line
spectrum. Therefore, photoionization, recombination and
collisions can be considered as relevant
processes in BLRs. At larger ionization parameters, recombination is
more important, but at the higher temperatures the collisional excitation
becomes also important as in the case of low ionization
parameters (Osterbrock 1989).
These two effects, together with the radiative-transfer effects in Balmer lines,
should be taken into account in explaining the
 ratios of Hydrogen lines. Moreover, the geometry and possible
 stratification in the BLR may also affect both continuum and line spectra
(Goad et al.
1993).}

Notice here that in investigation of the physical parameters of the BLR, first 
a theory is assumed (i.e. model of excitation, absorption of radiation, energy 
output from the center, etc.) and thereafter a comparison between the
observed and the predicted line ratios is discussed. There is a problem to 
develop a theory that can be applied to observations, i.e. to use the measured 
line ratios or line profiles to directly conclude about the physical 
parameters or physical conditions in the BLR. 

To conclude about physics in the BLR it is very useful to compare 
emission-line flux ratios with photoionization models. But also recently two 
methods are given that can be also used in determination (indication) of 
physical parameters in the BLR: (i) BP method given by Popovi\'c (2003, 2006) 
using the Balmer lines flux ratios and atomic parameters of Balmer lines, 
and (ii) electron scattering influence on line shapes given by Laor (2006) 
where the physical parameters can be determined by fitting the line wings. 

\subsection{3.1 Photoionization models}

First approach to use the photoinoization model started from assumption of the 
existence of a single cloud pressure law with radius (e.g. Rees et al. 1989).
The simple one-zone 
photoionization
models cannot properly describe BLRs since one can expect that gas clouds
in BLRs embrace wide ranges in densities and/or ionization degrees
in general (see e.g. Collin-Souffrin et al.
1982). In order to investigate the physical properties of gas clouds in
the BLR, Baldwin et al. (1995) proposed the Locally
Optimally emitting Cloud (LOC) model, that is a multizone
photoionization model. In this model, gas clouds with a
wide range of physical conditions are present at a wide range
of distances, and thus the net emission-line spectra can be calculated
by integrating in the parameter space of gas density
and radius, assuming different distribution functions. Using this model
one can calculate fluxes of both low-ionization emission lines and
high-ionization emission lines consistently and simultaneously.
It can be  used to investigate physical
 properties of ionized gas clouds in the BLR (e.g.
Korista and Goad 2004). Using the method, one  can obtain the
net emission-line flux by integrating the line emissivity of all
clouds

$$L_\mathrm{line} =\int\int 4\pi r^2F_\mathrm{line}(r,n)f(r)g(n)dndr\, ,$$
where $f(r)$ and $g(r)$ are the cloud distribution function and gas density, 
 respectively. The radius of the BLR is specified by the ionizing photon flux. 
Baldwin et 
 al (1995) assumed simple power-law functions for both $f(r)\sim r^\Gamma$ and 
$g(r)\sim 
 n^\beta$. It is shown that the BLR emission line spectra may be well 
reproduced by 
 the LOC models with $\Gamma\approx -1$ and $\beta \approx -1$ (Korista and Goad 
2000). The problem with model is, as mentioned above,  that in some cases 
photoionization models cannot explain well the flux ratios in some BELs (broad 
hydrogen lines as well as Fe II lines).

\subsection{3.2 Using the Balmer line ratios: the BP method}

Recently, Popovi\'c (2003, 2006) showed that in the
BLR of some AGN, the Balmer line ratios follow the Boltzmann-plot (BP). 

If we assume that plasma of the length $\ell$ along the line of sight emits, 
the flux (or the spectrally  integrated emission-line intensity ($I_{lu}$)) 
can be calculated as: 

$$I_{lu}={hc\over\lambda}g_{u}A_{ul}\int_0^\ell
N_u(x)dx\eqno(1)$$
where $\lambda$ is transition wavelength, $g_u$  statistical weight of
the upper level, $A_{ul}$  transition probability, $N_u$ is the number of emitters 
excited in upper level which effectively contribute to the line flux (which are not 
absorbed) and $h$ and  $c$ are the well
known constants (Planck and speed of light). In principle, the $N_u$ can be 
inhomogeneous across the line of 
sight and, also, the radiative self-absorption can be present. But, assuming that 
populations in the observed region (in all layers) follow the Boltzmann-Saha 
distribution one can 
write
$$N_u(x)\approx {N_0 (x)\over
Z}\exp(-E_u/kT_e(x)),\eqno(2)$$
where $Z$ is  the partition
function, $N_0$   the total number density  of radiating species,
$E_{u}$  the energy of the
upper level, $T_e$  electron temperature and $k$  the Boltzmann
constant.

In the case of optically thin plasma with relatively small variations in  
electron density and temperature, one can write (see e.g. Griem 1997, 
Konjevi\'c 1999) 

$$I_{lu}={hc\over\lambda}g_{u}A_{ul}\int_0^\ell
N_udx$$

\centerline{\includegraphics[height=8cm,angle=-90]{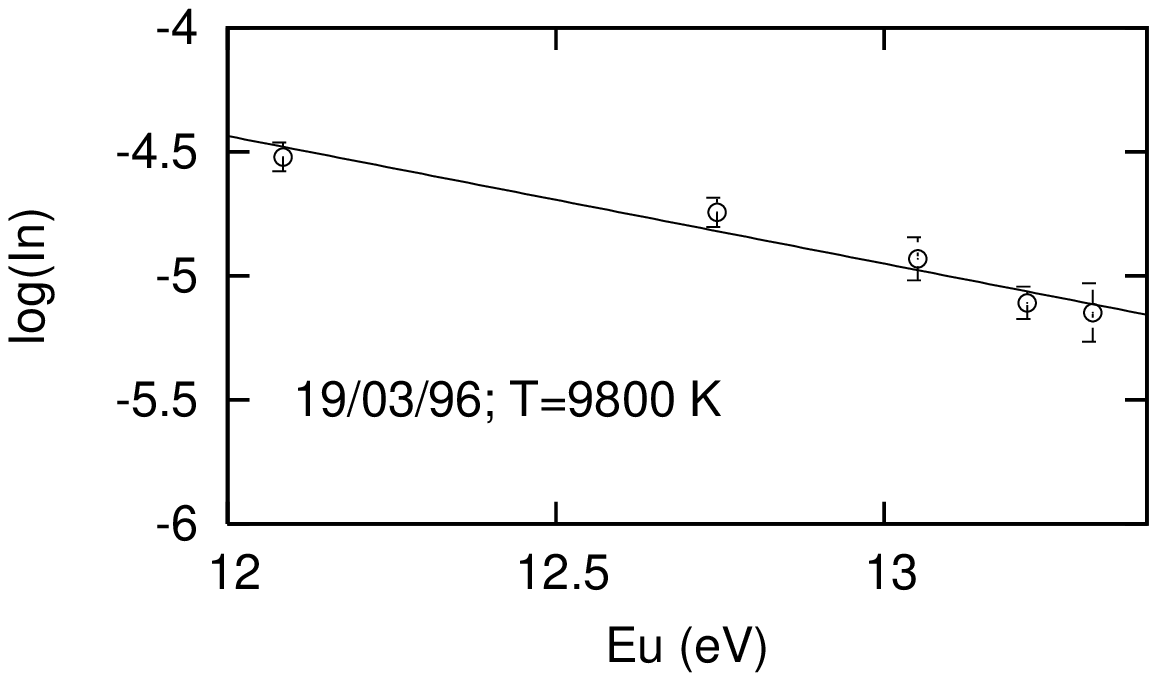}}
\centerline{\includegraphics[height=8cm,angle=-90]{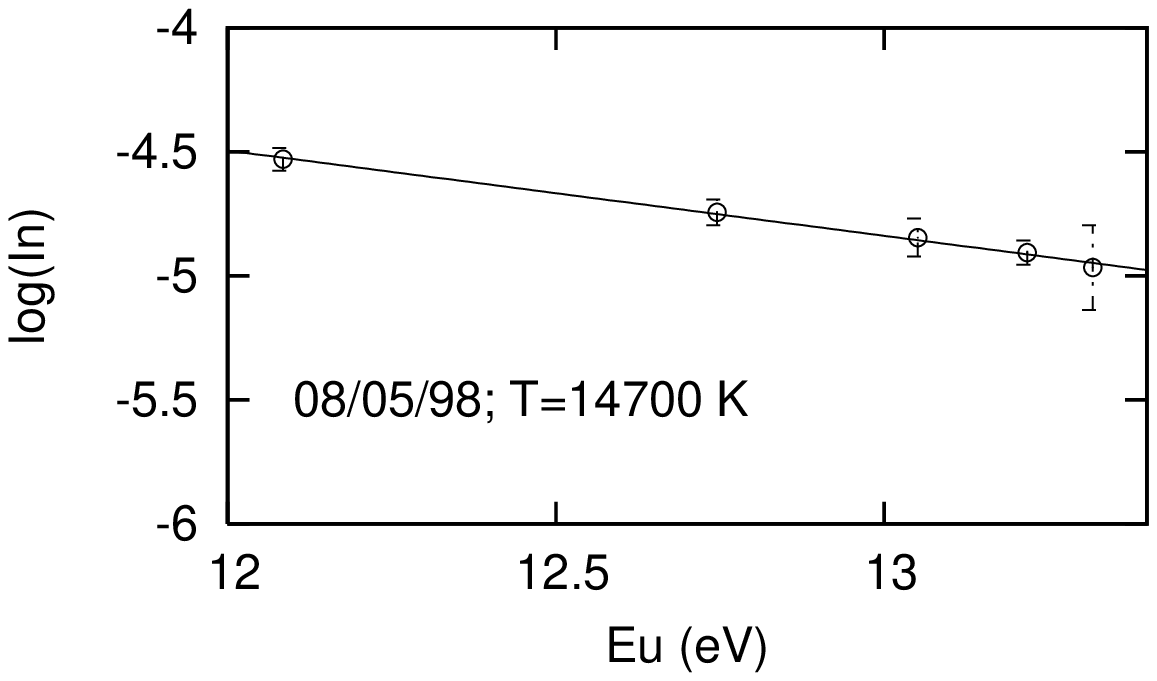}}
\centerline{\includegraphics[height=8cm,angle=-90]{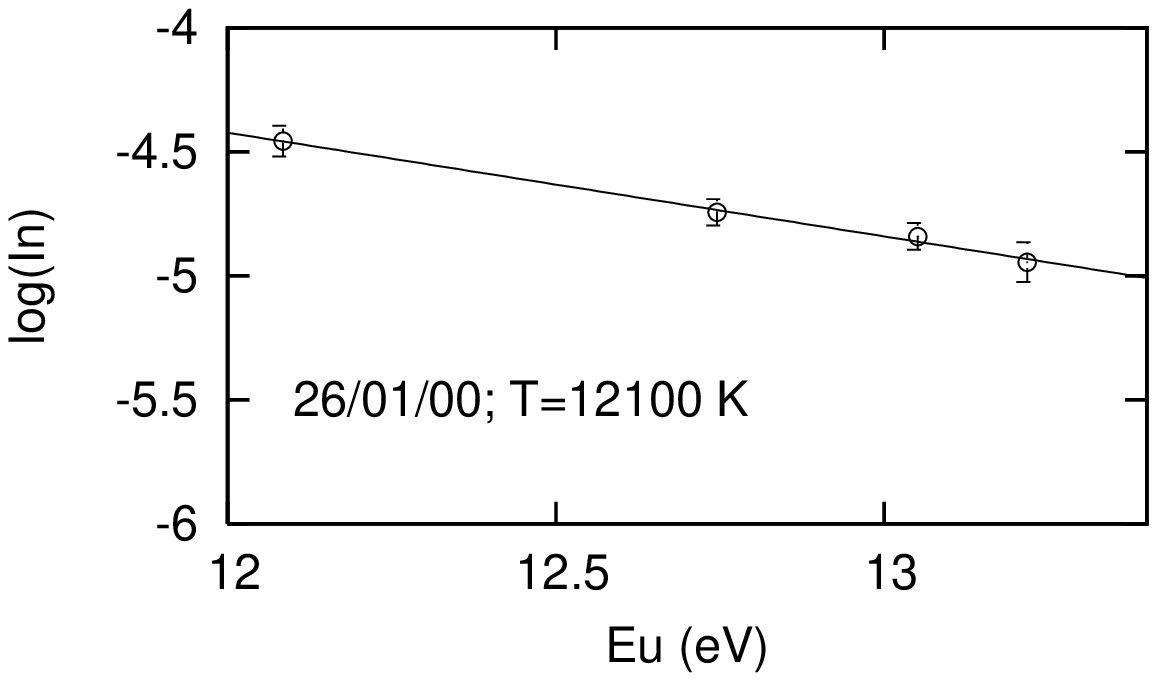}}

\figurecaption{6.}{The Boltzmann Plots of the Balmer line series of NGC 5548 from
different
periods (Popovi\'c et al. 2006c).}

$$\approx {hc\over\lambda}A_{ul}g_u\ell{N_0\over
Z}\exp(-E_u/kT_e),\eqno(3)$$
where $T_e$ is electron temperature.  For one line  series (e.g. the Balmer 
line series), if the population of the upper energy  states 
($n\ge3$)\footnote{Note here that it is because, in the emission, deexcitation 
goes as $u\to l$, 
it is not necessary that the level $l$ has a Boltzmann distribution.}  adhere to 
a Boltzmann distribution that one can determine their excitation temperature 
($T_e$) that can be obtained from a Boltzmann plot as 

$$\log(I_n)=\log{F_{ul}\cdot
\lambda\over{g_uA_{ul}}}=B-A{E_u},\eqno(4)$$
where $F_{lu}$ is the relative flux of a transition from upper to lower
level ($u\to l$), $B$ and $A$ are constants, where $A=1/kT$ is the
temperature indicator,\footnote{if we use $\log_{10}$ in Eq. (2), then
$A'=\log_{10}(e)/kT$, and this value will be used in the paper.} then from this
value we can estimate the electron
temperature from Eq. (4).

If all lines of a series follow the Boltzmann plot, as it is assumed in Eq. 
(4), from the constant $A$ the electron temperature can be determined, 
and we can say that for this series the Partial Thermodynamical Equilibrium 
(PLTE) exists (see van der Mullen et al. 1994, Fujimoto and McWhirter 1990). 
Note here, that the PLTE condition are not necessarily identical with the 
pure LTE 
conditions. The plasma in which PLTE exist for one series can be non-stationary 
and/or two-temperature plasma (for more details see van der Mullen et al. 
1994). Also, in investigation of the BLR physics, very often the Balmer line 
ratios or Balmer decrements are used. At first glance, the Boltzmann plot looks 
like the Balmer decrement, but it should be pointed out that it is NOT the 
same as the Balmer decrement, and the initial assumption in the Boltzmann plot 
is that the populations of the upper levels in e.g. Balmer series follow the 
Boltzmann-Saha equation. 

In the case of the PLTE, the BP method can be used for the electron 
temperature  diagnosting, {and vice versa, if it is possible to apply BP on 
a line series, it indicates the presence of the PLTE}. But, as it was already 
noted 
(Popovi\'c 2003), an alternative to PLTE in Balmer series of the BLR of some 
AGN might be  intrinsic reddening, and a value of $A>0.3$ is estimated to be 
the lower 
limit for $A$ that can be used in temperature determination. Note here, also, 
that the accuracy of this method depends on the number of lines available and 
on the energy interval ($\Delta E$) between excited levels involved, 
and that it should be $\Delta E>kT$.\footnote{For the Balmer series, $\Delta 
E\approx 1.23$ eV, which corresponds to $T\approx 15000$ K.} In the case $\Delta 
E<kT$, even if the BP works, the determined temperatures are with higher 
errors. In a prevous paper (Popovi\'c 2003) it is estimated that the accuracy 
of the determined temperature of BLRs by this method lies within the frame of 
30\%.  

On the other hand, one cannot expect the presence of a homogeneous distribution of physical 
parameters and density of emitters along the line of sight. But, if we still 
have the population following the Boltzmann-Saha equation, Eq. (1) can be 
written as: 

$$I_{lu}={hc\over\lambda}g_{u}A_{ul}\int_0^\ell {N_0 (x)\over
Z}\exp(-E_u/kT_e(x)) dx\eqno(5)$$
and if the BLR is divided in small layers with the same physical conditions and 
emitter density, we can write:

$$I_{lu}={hc\over\lambda}g_{u}A_{ul}\sum_{n=1}^n {N_0 (i)\over
Z}\exp(-E_u/kT_e(i)) \ell_i.\eqno(6)$$
Assuming that the temperatures across the BLR vary as $T_i=T_{av}\pm \Delta T_i$ 
and the emitter 
densities 
as $N_0(i)=N_0^{av}\pm\Delta N_0(i)$, Eq. (6) can be written as:

$$I_{lu}={hc\over\lambda}g_{u}A_{ul}{N_0^{av}\over
Z}\exp(-E_u/kT_{av}) \ell \times \delta(N_0,T)\eqno(7)$$
where
$$ \delta(N_0,T)=  \sum_1^n {{(1+\delta N_0 (i))}}\exp[1/(1+ \delta T_i) 
]\frac{\ell_i}{\ell},\eqno(8)$$
where $\delta T_i=\Delta T_i/T_{av}$ and $\delta N_0=\Delta N_0/N^{av}_0$. If 
in a BLR the  temperature and emitter density do not vary very much, i.e. if
$\Delta N_0/N_0<<1$ and $\Delta T_i/T_{av}<<1$, then $\delta(N_0,T)\approx 1$ 
and the Eq. (4) can be used to determine $T_{av}$ in the BLR. 

{Note here, that the self-absorption can also affect the Balmer line ratios (see 
e.g. Netzer 1975), but it seems that the BLR (or, at least, a part of the 
BLR)   is optically thin (e.g.  Corbin and Borson 1996, see also the discussion 
above). Moreover, the majority of broad lines in  AGN have no strong  asymmetry 
which should be present if self-absorption were dominant (Ferland et al. 
1979); 
of  course it does not automatically mean the absence of Balmer self-absorption 
(see Ferland et al. 1979), but it can also indicate that most of the observed 
H$\alpha$ emission results from  collisional excitation (Ferland et al. 1979). 
It seems that collisional excitation has leading role in formation of the 
Balmer lines in some BLRs (around 30\% of broad line AGN from a SDSS sample 
show that the Balmer lines follow BP, see La Mura et al. 2006). In this case, 
the opacity is proportional to the density of atoms in $n=2$ state and the 
collisional photon  generation rate for the Balmer lines will be distributed 
fairly uniformly throughout their optical depth, provided that the 
electron density and temperature do not vary drastically across  (one 
part of) the BLR, where  the collisional excitation is dominant. Then one can 
use the BP method for diagnosing an averaged temperature in the region. Such 
a scenario was proposed by Ferland et al. (1979) in order to explain the 
absence of line asymmetry in most of AGN. Moreover, this is consistent with 
the mentioned ratio of the Ly$\alpha$ and H$\alpha$ fluxes. Regarding this 
investigation, one can conclude that the largest part of the emitted photons 
in the broad Balmer lines (which follow the BP) are collisionally created. 

As noted in 2.1, the multicomponent BLR should have various 
physical properties and in this case one still can conclude about the physics 
using the BP; for more details see Popovi\'c et al. (2006c) and Ili\'c et al. 
(2006). 

Note here that the BP method  (Popovi\'c 2003, 2006) does not take into 
account any 'a priori' given physics in the BLR, besides   that Balmer lines 
originate in the same  emitting region. Also, the method includes the 
intensities of all lines from Balmer series, not only the ratio of some of the 
lines. Also, it should be noted here that at around of 30\% of AGN, the Balmer 
lines follow the BP (see La Mura et al. 2006). 

\subsection{3.3 The line profiles: Electron scattering}

The line profiles of broad lines emitted from the BLR are mainly affected by 
kinematics in the BLR. Recently, Laor (2006) showed that, in the low luminous 
AGN NGC 4395, the line shape can be affected by electron scattering. He used 
this to develop a method that uses the line profiles to determine the electron 
density and temperature.  The line profile is so-called logarithmic, and 
fitting the line wings one can obtain the physical parameters in the BLR,  
assumed to be optically thin (for more details, see Laor 2006). The problem 
with this method is that many effects can be expected to affect the line 
profiles (even the gravitational field, see e.g. 
Popovi\'c et al. 1995), and the method can be very useful in the case of lowly 
luminous AGN, as it was originally applied by Laor (2006). 

\vskip-2mm

\section{4. CONCLUSION}

\vskip-1mm

In this paper, the geometry and physical properties in the BLR are discussed. 
The BLR is very close to the massive black hole and is affected by strong 
radiation from the central part. Various geometries can be considered in 
the attempts to explain the complex broad line shapes of AGN. Probably, the 
BLR is complex and may be composed of two or more geometrically different 
regions. But, according to a standard model, one can expect that an accretion 
disk emission is  present in the BLR. This emission should mainly contribute 
to the broad line wing shapes. On the other hand, the electron temperature and 
density should also vary across the BLR, but it seems that, in a number of 
AGN, the population of the Balmer series follow Saha-Boltzmann equation and  
one can at least have information about an average temperature in the region. 
This can be of help in applying more sophisticated methods for calculation of 
the physical conditions in the BLR. 
}

\vspace{-2mm}


\acknowledgements{This work is a part of the project P146002  "Astrophysical 
Spectroscopy of
Extragalactic Objects" supported by the
Ministry of Science and Environment Protection
of Serbia.}

\vskip.6cm


\references

\vspace{-2mm}

Abajas, C., Mediavilla, E., Mu\~noz, J. A., Popovi\'c, L. \v C.: 2005, 
\journal{Mem. Soc. Astron. Ital. Supp.}, \vol{7}, 48. 

Abajas, C., Mediavilla, E., Mu\~noz, J. A., Popovi\'c, L. \v C. Oscoz, A.: 
2002, \journal{Astrophys. J.}, \vol{565}, 105. 

Antonucci, R.: 1993, \journal{Annu. Rev. Astron. Astrophys.}, \vol{31}, 473.


Baldwin, J., Ferland, G., Korista, K.,  Carswell, R.F.
Hamann, F., Phillips, M.M.,  Verner, D., Wilkes, B.J. and Williams R.E.:
1996, \journal{Astrophys. J.}, \vol{461}, 664.

Baldwin, J.,  Ferland, G.,  Korista, K.,  Verner, D.: 1995, 
\journal{Astrophys. J.}, \vol{455L}, 119.


Bon, E., Popovi\'c, L. \v C., Ili\'c, D., Mediavilla, E.: 2006, \journal{New 
Astron. Rev.}, \vol{50}, 716. 

Chen, K., Halpern, J.P. and Filippenko, A.V.: 1989, \journal{Astrophys. J.}, 
\vol{339}, 742. 

Chen, K. and Halpern, J.P.: 1989, \journal{Astrophys. J.}, \vol{344}, 115.

Collin-Souffrin, S.: 1986, \journal{Astron. Astrophys.}, \vol{166}, 115.

Collin-Souffrin, S., Dumont, S.: 1986, \journal{Astron. Astrophys.}, 
\vol{166}, 13. 

Collin-Souffrin, S., Dumont, S., Tully, J.: 1982, \journal{Astron. 
Astrophys.}, \vol{106}, 362.

Corbin, M.R. and Boroson, T.A.: 1996, \journal{Astrophys. J. Suppl. Series}, 
\vol{107}, 69.

Crenshaw, D.M., Kraemer, S.B., Gabel, J.R., Schmitt, H.R.,
Filippenko, A.V., Ho, L.C., Shields, J.C., Turner, T.J.: 2004, 
\journal{Astrophys. J.}, \vol{612}, 152.

Crenshaw, D.M., Kraemer, S.B., Turner, T.J. et al.: 2002,
\journal{Astrophys. J.}, \vol{566}, 187.


Dimitrijevi\'c, M.S., Popovi\'c, L.\v C., Kova\v cevi\'c, J.,
Da\v ci\'c, M., Ili\'c, D.: 2006, \journal{Mon. Not. R. Astron. Soc.}, accepted 
(astro-ph/0610848)

Dultzin-Hacyan, D., Marziani, P., Sulentic, J.W.: 2000, \journal{Rev. Mex. 
Astron. Astrophys.}, 
\vol{9}, 308. 

Dumont, A.M., Collin-Souffrin, S.: 1990a, \journal{Astron. Astrophys.}, 
\vol{229}, 292.

Dumont, A.M. and Collin-Souffrin, S.: 1990b, \journal{Astron. Astrophys. 
Suppl. Series}, \vol{83}, 71.

Dumont, A.-M., Collin-Souffrin, S., Nazarova, L.: 1998, \journal{Astron. 
Astrophys.}, \vol{331}, 11.

Dumont, A.M. and Joly, M.: 1992, \journal{Astron. Astrophys.}, \vol{263}, 75.

Dumont, A.M., Lasota, J.P., Collin-Souffrin, S., King, A.R.: 1991, 
\journal{Astron. Astrophys.}, \vol{242}, 503.

Gabel, J.R., Kraemer, S.B., Crenshaw, D.M. et al.:
2005, \journal{Astrophys. J.}, \vol{631}, 741.

Gaskell, C.M.: 1982, \journal{Astrophys. J.}, \vol{263}, 79.

Gaskell, C.M.: 1996, \journal{Astrophys. J. Lett.}, \vol{464}, 107.

Goad, M.R., O'Brien, P.T. and Gondhalekar, P.M.: 1993, \journal{Mon. Not. R. 
Astron. Soc.}, \vol{263}, 149.

Goad, M. and Wanders, I.: 1996, \journal{Astrophys. J.}, \vol{469}, 113.

G\'omez-\'Alvarez, P., Mediavilla, E., Mu\~noz, J. A., Arribas, S., S\'anchez, 
S. F., Oscoz, A., Prada, F., Serra-Ricart, M.: 2006, \journal{Astrophys. J.}, 
\vol{645}, 5. 

Griem, H.R.: 1997, Principles of Plasma Spectroscopy, Cambridge
University Press.

Eigenbrod, A., Courbin, F., Dye, S., Meylan, G., Sluse, D., Vuissoz, C., 
Magain, P.: 2006, \journal{Astron. Astrophys.}, \vol{451}, 747. 

Elvis, M.: 2000, \journal{Astrophys. J.}, \vol{545}, 63.

Eracleous, M. and Halpern, J.P.: 1994, \journal{Astrophys. J. Suppl. Series}, 
\vol{90}, 1.

Eracleous, M. and Halpern, J.P.: 2003, \journal{Astrophys. J.}, \vol{599}, 886.

Ferland, G.J., Korista, K.T., Verner, D.A., Ferguson, J.W., Kingdon, J.B., 
Verner, E.M.: 1998, \journal{Publ. Astron. Soc. Pacific}, \vol{110}, 761. 

Ferland, G.J. and Netzer, H.: 1979, \journal{Astrophys. J.}, \vol{229}, 274.

Ferland, G. J., Netzer, H., and Shields, G. A. 1979, \journal{Astrophys. J.}, 
\vol{232}, 382.

Ferland, G.J., Peterson, B.M., Horne, K., Welsh, W.F., Nahar, S.N.: 1992, 
\journal{Astrophys. J.}, \vol{387}, 95.

\endreferences

\end{multicols}

\vfill\eject

\begin{multicols}{2}

\refcontinue



Fujimoto, T. and McWhirter, R.W.P.: 1990, \journal{Phys. Rev. A}, \vol{42}, 
6588. 

Halpern, J.P., Eracleous, M., Filippenko, A.V., Chen, K.: 1996, 
\journal{Astrophys. J.}, \vol{464}, 704. 


Ili\'c, D.,  Popovi\'c, L.\v C., Bon, E., Mediavilla, E.G., Chavushyan, V.H.: 
2006, \journal{Mon. Not. R. Astron. Soc.}, \vol{371}, 1610.

Joly, M.: 1987, \journal{Astron. Astrophys.}, \vol{184}, 33.

Joly, M.: 1988, \journal{Astron. Astrophys.}, \vol{192}, 87.

Joly, M.: 1991, \journal{Astron. Astrophys.}, \vol{242}, 49.

Jovanovi\'c, P.: 2006, \journal{Publ. Astron. Soc. Pacific}, \vol{118}, 656.

Kallman T., Krolik J., 1986, \journal{Astrophys. J.}, \vol{308}, 80.

Kaspi, S., Smith, P.S., Netzer, H., Maoz, D., Jannuzi, B.T., Giveon, 
U.: 2000, \journal{Astrophys. J.}, \vol{533}, 631.

Kollatschny, W. and Bischoff, K.: 2002, \journal{Astron. Astrophys.}, 
\vol{386}, L19.

Kollatschny, W.: 2003, \journal{Astron. Astrophys.}, \vol{407}, 461.

Konjevi\'c, N.: 1999, \journal{Physics Reports}, \vol{316}, 339.



Korista, K.T. and Goad, M.R.: 2000, \journal{Astrophys. J.}, \vol{536}, 284.

Korista, K.T. and Goad, M. R. 2004, \journal{Astrophys. J.}, \vol{606}, 749.


Krolik, J.: 1999, Active Galactic Nuclei: From the Central Black Hole to
the Galactic Environment (Princeton: Princeton Univ. Press).

Kwan, J.: 1984, \journal{Astrophys. J.}, \vol{283}, 70.

Kwan, J. and Krolik, J.H.: 1981, \journal{Astrophys. J.}, \vol{250}, 478.

La Mura, G., Popovi\'c, L.\v C., Ciroi, S., Rafanelli, P., Ili\'c, D.: 2006, 
\journal{Astrophys. J.}, submitted. 

Lamer, G., Schwope, A., Wisotzki, L., Christensen, L.: 2006, \journal{Astron. 
Astrophys.}, \vol{454}, 493. 

Laor, A.: 2006, \journal{Astrophys. J.}, \vol{643}, 112.

L\'ipari, S.L., Roberto and Terlevich, R.: 2006, \journal{Mon. Not. R. Astron. 
Soc.}, \vol{368}, 1001. 


Marziani, P., Sulentic, J.W., Dultzin-Hacyan, D., Calvani, M., Moles, M.: 
1996, \journal{Astrophys. J. Suppl. Series}, \vol{104}, 37.

Miller, J.S., Goodrich, R.W.: 1990, \journal{Astrophys. J.}, \vol{355}, 456.

Murray, N. and Chiang, J.: 1997, \journal{Astrophys. J.}, \vol{474}, 91.


Netzer, H.: 1975, \journal{Mon. Not. R. Astron. Soc.}, \vol{171}, 395.

Netzer, H.: 1976, \journal{Mon. Not. R. Astron. Soc.}, \vol{177}, 473.

Norman, C., Miley, G.: 1984, \journal{Astron. Astrophys.}, \vol{141}, 85.

Osterbrock, D.E.: 1989, Astrophysics of Gaseous Nebulae and Active
Galactic Nuclei.

Pariev, V.I. and Bromley, B.C.: 1998, \journal{Astrophys. J.}, {\vol 508}, 
590.

Perez, E., Mediavilla, E., Penston, M.V., Tadhunter, C., Moles, M.: 1988, 
\journal{Mon. Not. R. Astron. Soc.}, \vol{230}, 353. 

Peterson, B.M.: 2003, An Introduction to Active Galactic Nuclei, Cambridge 
University Press.

Peterson, B.M.: 1993, \journal{Publ. Astron. Soc. Pacific}, \vol{105}, 247.

Peterson, B.M., Barth, A.J., Berlind, P. et al.: 1999, \journal{Astrophys. 
J.}, \vol{510}, 659. 

Popovi\'c, L.\v C.: 2003, \journal{Astrophys. J.}, \vol{599}, 140.

Popovi\'c, L.\v C.: 2005, Proc. of the ESO Astrophysics Symposia 
"Science Perspectives for 3D Spectroscopy" 
(eds, M. Kissler-Patig, M.M. Roth and J.R. Walsh), Garching, 
Germany, 10-14 October 2005 (astro-ph/0512594)

Popovi\'c, L.\v C. 2006, \journal{Astrophys. J.}, \vol{650}, 1217.

Popovi\'c, L.\v C. and Chartas, G.: 2005, \journal{Mon. Not. R. Astron. Soc.}, 
{\vol 357}, 135. 

Popovi\'c, L.\v C., Mediavilla, E., Bon, E., Ili\'c, D.: 2004, 
\journal{Astron. Astrophys.}, \vol{423}, 909. 

Popovi\'c, L.\v C., Mediavilla, E.G., Mu\~noz, J.A., Dimitrijevi\'c, M.S., 
Jovanovi\'c, P.: 2001b, \journal{Serb. Astron. J.}, \vol{164}, 53.

Popovi\'c, L.\v C., Mediavlilla, E.G., Bon, E., Stani\'c, N.,
Kubi\v cela, A.: 2003a, \journal{Astrophys. J.}, \vol{599}, 185.

Popovi\'c, L.\v C., Mediavilla, E.G., Pavlovi\'c, R.: 2000, \journal{Serb. 
Astron. J.}, \vol{162}, 1. 

Popovi\'c, L.\v C., Mediavilla, E.G., Jovanovi\'c, P.,  Mu\~noz, J.A.:
 2003b,  \journal{Astron. Astrophys.}, \vol{398}, 975.

Popovi\'c, L.\v C., Jovanovi\'c, P., Mediavilla, E.G.,  Mu\~noz,
J.A.: 2003c, \journal{Astron.  Astrophys. Transactions}, \vol{22}, 719.

Popovi\'c, L.\v C., Jovanovi\'c, P.,  Mediavilla, E., Zakharov, A.F., Abajas,
C., Muoz, J.A., Chartas, G.: 2006a, \journal{Astrophys. J.}, \vol{637}, 620.

Popovi\'c, L.\v C., Jovanovi\'c, P.,  Petrovi\'c, T., Shalyapin, V.N.: 2006b, 
\journal{Astron. Nachr.}, {\vol 327}, 981.

Popovi\'c, L.\v C., Mediavlilla, E.G., Kubi\v cela, A.,
Jovanovi\'c, P.: 2002, \journal{Astron. Astrophys.}, \vol{390}, 473.

Popovi{\'c}, L.\v C., Mediavilla, E.G., Mu\~noz, J.: 2001b,
\journal{Astron. Astrophys.}, \vol{378}, 295.

Popovi\'c, L.\v C., Stani\'c, N., Kubi\v cela, A., Bon, E.:
2001a, \journal{Astron. Astrophys.}, \vol{367}, 780.

Popovi\'c, L.\v C., Shapovalova, A.I., Chavushyan, V.H., Ilic, D., Burenkov, 
A.N., Mercado, A.: 2006c, astro-ph/0511676

Popovi\'c, L.\v C., Vince, I., Atanackovi\'c-Vukmanovi\'c, O., Kubi\v cela, 
A.: 1995, \journal{Astron. Astrophys.}, \vol{293}, 309.

Rees, M.J., Netzer, H. and Ferland, G.J. 1989, \journal{Astrophys. J.}, 
\vol{347}, 640.

Reichard, Timothy A.; Richards, Gordon T.; Hall, Patrick B.;
Schneider, Donald P.; Vanden Berk, Daniel E.; Fan, Xiaohui; York,
Donald G.; Knapp, G. R.; Brinkmann, J.: 2003, \journal{Astron. J.}, \vol{126}, 2594.

Richards, G.T., Keeton, C.R., Pindor, B. et al.: 2004, \journal{Astrophys. 
J.}, \vol{610}, 679. 


Rokaki, E. and Boisson, C.: 1999, \journal{Mon. Not. R. Astron. Soc.}, 
\vol{307}, 41.

Romano, P., Zwitter, T., Calvani, M., Sulentic, J.: 1996, \journal{Mon. Not. 
R. Astron. Soc.}, \vol{279}, 165. 

Shapovalova, A.I., Doroshenko, V.T., Bochkarev, N.G. et al.: 
2004, \journal{Astron. Astrophys.}, \vol{422}, 925.

Shields, G.: 1977, \journal{Astrophys. Lett.}, \vol{18}, 119.

Shields, J.C. and Freland, G.J.: 1993, \journal{Astrophys. J.}, \vol{402}, 
425.

Sigut, T.A.A. and Pradhan, A.K.: 2003, \journal{Astrophys. J. Suppl. Series}, \vol{145}, 15.

Smith, J.E., Robinson, A., Young, S., Axon, D.J., Corbett, E.A.: 2005 
\journal{Mon. Not. R. Astron. Soc.}, \vol{359}, 846. 



Strateva, I.V., Strauss, M.A., Hao, L. et al.: 2003, \journal{Astron. J.}, \vol{126}, 
1720. 

\endreferences

\end{multicols}

\vfill\eject

\begin{multicols}{2}

\refcontinue

Sulentic, J.W., Marziani, P., Zwitter, T., Calvani, M. and Dultzin-Hacyan,
D.: 1998, \journal{Astrophys. J.}, \vol{501}, 54.

Sulentic, J.W., Marziani, P. and Dultzin-Hacyan, D.: 2000,
\journal{Annu. Rev. Astron. Astrophys.}, \vol{38}, 521.


van der Mullen, J.A.M., Benoy, D.A., Fey, F.H.A.G., van der Sijde, B., Vi\v 
cek, J.: 1994, \journal{Phys. Rev. E}, \vol{50}, 3925. 

{\ }

V\'eron-Cetty, M.-P., Joly, M., V\'eron, P., Boroson, T., Lipari, S., Ogle, 
P.: 2006, \journal{Astron. Astrophys.}, \vol{451}, 851.

Wang, J.-M. Ho, L.C., Staubert, R.: 2003, \journal{Astron. Astrophys.}, 
\vol{409}, 887.

Zheng, W., Binette, L., Sulentic, J.W.: 1990, \journal{Astrophys. J.}, 
\vol{365}, 115.

Zheng, W., Veilleux, S., Grandi, S.A.: 1991, \journal{Astrophys. J.}, 
\vol{381}, 418.

\endreferences

\end{multicols}

{\ }


{\ }



\naslov{XIROKOLINIJSKI REGION KOD AKTIVNIH GALAKTIQKIH JEZGARA:\break 
KINEMATIKA I FIZIKA} 


\authors{L. \v C. Popovi\'c}

\vskip3mm


\address{Astronomical Observatory, Volgina 7, 11160 Belgrade 74, Serbia}

\vskip.7cm


\centerline{UDK \udc}

\vskip1mm


\centerline{\rit Pregledni rad po pozivu}

\vskip.7cm

\begin{multicols}{2}
{


\rrm U radu je data diskusija o fizici i kinematici u xirokolinijskoj oblasti 
kod aktivnih galaktiqkih jezgara. Razmatraju se mogu{\cc}i fiziqki uslovi u  
ovoj oblasti i diskutuju problemi vezani za odredjivanje fiziqkih parametara. 
Tako{\dd}e se razmatra mogu{\cc}nost da bar jedan deo emisije u xirokim 
linijama dolazi iz akrecionog diska. 
} 

\end{multicols} 

\end{document}